\documentclass[sigconf,natbib=true,screen]{acmart}
\usepackage{amsmath,amsfonts}
\usepackage{algorithmic}
\usepackage{graphicx}
\usepackage{textcomp}
\usepackage{xcolor}
\usepackage{colortbl}  
\usepackage{array}
\usepackage{url}
\usepackage{hyperref}
\usepackage[linewidth=1pt]{mdframed}
\usepackage{lipsum}
\usepackage{paralist}
\usepackage{listings}
\usepackage{courier}
\usepackage{xspace}
\usepackage{algorithm} 
\usepackage{algorithmic}
\usepackage{multirow}
\usepackage{framed}
\usepackage[]{graphicx}
\usepackage{caption}
\usepackage{subcaption}
\usepackage{enumerate}
\usepackage{enumitem}
\usepackage{tcolorbox}
\usepackage{lipsum} 
\usepackage{xcolor}
\usepackage[]{mdframed}
\usepackage{balance}
\usepackage{microtype}

\AtBeginDocument{%
  \providecommand\BibTeX{{%
    \normalfont B\kern-0.5em{\scshape i\kern-0.25em b}\kern-0.8em\TeX}}}

\acmConference[ESEC/FSE '23]{Proceedings of the 31st ACM Joint European Software Engineering Conference and Symposium on the Foundations of Software Engineering}{December 3--9, 2023}{San Francisco, CA, USA}
\acmBooktitle{Proceedings of the 31st ACM Joint European Software Engineering Conference and Symposium on the Foundations of Software Engineering (ESEC/FSE '23), December 3--9, 2023, San Francisco, CA, USA}
\received{2023-02-02}
\received[accepted]{2023-07-27}
\begin{document}

\title{Understanding Solidity Event Logging Practices in the Wild}
\titlenote{Zhongxing Yu is the corresponding author of this work, and the first three authors (Lantian Li, Yejian Liang, and Zhihao Liu) contributed equally to this work.}

\author{Lantian Li}
\affiliation{%
  \institution{Shandong University}
  \city{}
  \state{}
  \country{}}
\email{lilantian@mail.sdu.edu.cn}

\author{Yejian Liang}
\affiliation{%
  \institution{Shandong University}
  \city{}
  \state{}
  \country{}}
\email{yejianliang@mail.sdu.edu.cn}

\author{Zhihao Liu}
\affiliation{%
  \institution{Shandong University}
  \city{}
  \state{}
  \country{}}
\email{zhihaoliu@mail.sdu.edu.cn}

\author{Zhongxing Yu}
\affiliation{%
  \institution{Shandong University}
  \city{}
  \state{}
  \country{}}
\email{zhongxing.yu@sdu.edu.cn}

\begin{abstract}
Writing logging messages is a well-established conventional programming practice, and it is of vital importance for a wide variety of software development activities. The logging mechanism in Solidity programming is enabled by the high-level event feature, but up to now there lacks study for understanding Solidity event logging practices in the wild. To fill this gap, we in this paper provide the first quantitative characteristic study of the current Solidity event logging practices using 2,915 popular Solidity projects hosted on GitHub. The study methodically explores the pervasiveness of event logging, the goodness of current event logging practices, and in particular the reasons for event logging code evolution, and delivers 8 original and important findings. The findings notably include the existence of a large percentage of independent event logging code modifications, and the underlying reasons for different categories of independent event logging code modifications are diverse (\emph{e.g.}, bug fixing and gas saving). We additionally give the implications of our findings, and these implications can enlighten developers, researchers, tool builders, and language designers to improve the event logging practices. To illustrate the potential benefits of our study, we develop a proof-of-concept checker on top of one of our findings and the checker effectively detects problematic event logging code that consumes extra gas in 35 popular GitHub projects and 9 project owners have already confirmed the detected issues. 
\end{abstract}

\ccsdesc[500]{Software and its engineering~Maintaining software}
\ccsdesc[500]{Software and its engineering~Language features}
\ccsdesc[500]{General and reference~Empirical studies}

\keywords{Solidity, Ethereum, event, logging, empirical study}

\maketitle

\vspace{-2.0mm}
\section{Introduction} 
Ethereum is widely recognized as a distributed single-state world computer, which innovatively combines the computing architecture of a typical, general-purpose stored-program computer with a decentralized blockchain \cite{mastering}. In particular, the Ethereum platform features the execution of arbitrary programs termed \emph{smart contracts}, which are registered immutably on the blockchain and have their correct executions enforced by the consensus protocol \cite{ideasmartcontract}. 
As smart contracts are programmable, decentralized, and transparent, they promise to renovate plenty of areas (\emph{e.g.}, financial institutes, supply chains, and government governance), and recent years have witnessed a snowballing application on these areas \cite{application1, application3, application4}. 

Smart contracts are essentially Turing-complete programs that run in a highly constrained and minimalistic execution environment named Ethereum virtual machine (EVM), which runs a particular kind of low-level programming language code called \emph{EVM bytecode}. While in principle any high-level programming language could be adjusted to write smart contracts, it is extremely cumbersome to adapt an arbitrary language for compliance with EVM bytecode. Consequently, smart contracts in practice are programmed by a few languages specifically designed for writing smart contracts, including for example \emph{Solidity}, \emph{Vyper}, \emph{Serpent}. Among these languages, Solidity is up to now the most popular and even widely perceived as the de facto language of Ethereum \cite{mastering}. 
Solidity features the \emph{imperative} programming paradigm, and its syntax resembles that of JavaScript, C++, or Java. A Solidity contract is first compiled into EVM bytecode and then executed by EVM on the blockchain. 

As smart contract issues can literally cost money, the contract code is extremely expensive, if not impossible, to change after deployment, and the whole program state of smart contracts is transparent to everyone, it is of vital importance to develop high-quality smart contract programs. However, the present quality of Solidity programs is far from satisfactory and the literature \cite{DAOattack,application3,typeanalysis,modelcheking} has reported a variety of issues about even deployed Solidity contracts. We argue that one fundamental reason for this bad status lies in that developers are not adequately acquainted with Solidity language features. Being a new programming language, Solidity offers some new, counter-intuitive high-level language features \cite{counterintuition} which are unfamiliar to developers who are even quite adept at programming with other languages of longer history. As an example, Solidity features exceptions like many other languages, but with a peculiar behavior. When an exception is thrown, it cannot be caught: the execution stops and all the side effects — including transfers of ether — are reverted. Security problems like \emph{gasless send} can arise because of this. Moreover, the formal specification of Solidity language features is absent and the current documentation about these features is ambiguous and incomplete. To ameliorate this issue, an effective solution is gathering empirical evidence about the \emph{practices} applied by Solidity developers when using language features, and using the evidence to impact at least four audiences: 

\vspace{1.0mm}
\noindent (1) \emph{Researchers} are informed about the real unsolved issues faced by the Solidity
developers, and hence set research agenda to advance the present state-of-the-art.

\vspace{0.5mm}
\noindent (2) \emph{Language designers} are aware of whether the language features they design are correctly used by the Solidity developers, or are instead misused or underused.

\vspace{0.5mm}
\noindent (3) \emph{Tool builders} realize how to customize their tools (\emph{e.g.}, coding assistants) to the real needs and practices of the Solidity developers.

\vspace{0.5mm}
\noindent (4) \emph{Developers} understand both the good and bad practices concerned with the use of Solidity language features, and consequently improve the quality of the written code.
\vspace{1.0mm}

Writing logging messages is a well-established programming practice \cite{logging}, and it plays an extremely important role in various software development activities \cite{SherLog,LogEhance,yuan2012conservative,fu2014developers,li2018studying,beschastnikh2011mining,he2021survey,li2019dlfinder,li2020qualitative}, including for example failure diagnosis, auditing, and profiling. The logging mechanism in Solidity programming is enabled by the high-level \emph{event} feature, which is an abstraction on top of the logging primitives of EVM. Compared to conventional logging, Solidity event logging features a few distinctions. First, event logging prints the logging message to Ethereum blockchain instead of the console or file routinely done by conventional logging. Second, while the target application scenario of conventional logging mainly revolves around failure diagnosis, the target application scenario of event logging is not well established and varies. Event logging in principle can have a wide variety of application scenarios, and the reported scenarios include for example alternative cheaper storage venues and debugging \cite{antonopoulos2018mastering}. Given these distinctions, it is of vital importance to study how Solidity developers use event logging in practice and thus positively impact researchers, language designers, tool builders, and developers to improve the event logging practices. 

To fill the gap, we in this paper conduct a large-scale empirical study about Solidity event logging practices in the wild. In particular, we aim to shed light on these three questions: 
\vspace{-0.1cm}
\begin{itemize}[leftmargin=*]
\item {\textbf{RQ1 (Pervasiveness of Event Logging): How common is event logging?} This question aims to explore whether event logging feature is indeed widely used by developers in practice.}

\item {\textbf{RQ2 (Goodness of Event Logging Practice): Is current event logging practice good enough ?} This question aims to study whether developers use event logging feature in a right way.}

\item {\textbf{RQ3 (Reasons for Event Logging Code Modifications): Why do developers modify event logging code?} This question aims to understand the specific reasons for developers' modifications to event logging code.}
\end{itemize}
\vspace{-0.1cm}

{To answer these questions, we choose 2,915 popular open source Solidity projects on GitHub as subjects. For \textbf{RQ1}, we explore the density of event logging instruction in source code. For \textbf{RQ2}, we first study the churn rate \cite{icse05codechurn} according to the code revision history and then separate those event logging code modifications that are truly modifying event logging code as after-thoughts from those that are solely consistency updates along with other non-event logging code changes. We hereafter term the former and latter modifications as \emph{independent} and \emph{dependent} event logging code modifications respectively. By ``after-thought'', we here emphasize that independent event logging code modifications arise because event logging code is not written right by developers at the first attempt, they later detect the problem and modify the event logging code accordingly. Unlike dependent event logging code modifications which are unlikely to reflect much developers’ concerns over the event logging code itself, independent event logging code modifications are prone to reflecting more directly developers’ concerns over the event logging code. We thus focus on independent event logging code modifications in this paper. For \textbf{RQ3}, we randomly sample 419 independent event use code modifications and divide them into different categories. Moreover, we investigate the details of them to understand the underlying reason for each category. }

{Our large-scale study enables us to deliver 8 original and important findings. In short, the findings include: (1) The average value of event logging code per project and per line of code (LOC) is 32.8 and 0.022
respectively, yet the maximal values can be exceptionally large; 
(2) The average churn rate of event logging code is nearly the same as that of the entire code, and a significant percentage (with lower bound being 10.64\%) of event logging code modifications are independent ones; (3) Within independent event logging code modifications, three major change categories include parameter change of event logging code, addition of new event logging code, and deletion of existing event logging code. The underlying reasons for these change categories are diverse, and notably are related with bug fixing, gas saving, and debugging support. We additionally give the implications of our findings, and these implications can enlighten developers, researchers, tool builders, and language designers to improve the practices of event logging.}

To illustrate the potential benefits brought by our study, we develop a proof-of-concept checker to detect problematic event logging code that consumes extra gas by using Storage type variable instead of Memory type variable (inspired by our Finding 5). For the top 200 popular GitHub Solidity projects, our checker detects that 35 projects suffer from this issue in the latest versions of their code (with 207 problematic event logging instructions in total). The owners of 9 projects have confirmed the detected problems and some of them commented that our finding is extremely interesting and can be integrated into the compiler optimization process. The result certifies that our findings are truly beneficial to tool builders for improving the quality of Solidity event logging instructions. 

This paper makes the following major contributions:
\begin{itemize}[leftmargin=*]
\item We conducted the first large-scale and systematic empirical study about Solidity event logging practices on 2,915 popular open-source Solidity projects hosted on GitHub.

\item We presented 8 original and important empirical findings about Solidity event logging practices. We additionally presented 8 implications from our findings, and these implications are beneficial for researchers, language designers, tool builders, and developers in order to improve all facets of Solidity event logging.

\item We developed a checker to detect problematic event logging code that consumes extra gas by using Storage type variable instead of Memory type variable, and the usefulness of the checker has been confirmed. 
\end{itemize}

\section{Background} 

\subsection{Ethereum Basics}
Unlike Bitcoin (with a quite constrained scripting language), Ethereum has been designed as a general-purpose, Turing-complete programmable blockchain that runs the EVM capable of executing smart contracts of arbitrary complexity \cite{antonopoulos2018mastering}. Smart contracts will execute transactions automatically in case certain conditions have been met, and can invoke other contracts during the execution of transactions initiated by external users.

Being a Turing-complete computing model, to ensure that denial-of-service attacks or transactions that consume excessive resources are avoided, Ethereum introduces a metering mechanism called gas to control the use of resources by transactions. More specifically, Gas measures the computational and storage resources required for different EVM instructions, and it has an exchange rate relationship with ether (the native Ethereum cryptocurrency). The required gases for different EVM instructions vary greatly, depending majorly on where the associated data reside. EVM can store data in three different places: Storage, Memory, and Stack. While most instructions cost 3-10 gas, Storage usage instructions like \texttt{sstore} can cost as much as 20,000 gas because Storage data are persistent on the blockchain (similar to hard drive of computer). To prevent malicious infinite loops and other forms of computational waste, the creator of each transaction needs to set a limit on the amount of gas that they are willing to pay for transaction execution. If the resource consumption during the execution exceeds the gas limit specified, the EVM will terminate the execution of the contract. 

Beyond the initial vision of being a general-purpose blockchain, the vision of Ethereum has rapidly expanded to become a platform for programming DApps (i.e., Decentralized
Applications). In a DApp, smart contracts serve as the backend and are employed to store the business logic (program code) and the related state of application, but the frontend can make use of standard web technologies, including typically HTML, CSS, and JavaScript.

\subsection{Solidity Event Definition and Use}
The message logging in Solidity code is achieved by the high-level event feature, which in turn will be compiled into EVM logging primitives. EVM has 5 logging primitives \texttt{log0}, \texttt{log1}, \texttt{log2}, \texttt{log3}, and \texttt{log4}, and these primitives can be used to create log records that describe an event within a smart contract. Each log record consists of two parts, \texttt{topic} and \texttt{data}. The topic part is used to describe what’s going on in an event, and the data part is the payload (value) of the event. The 5 EVM logging primitives differ in the number of topics that need to be included in the log record, and \texttt{logi} will include \texttt{i} topic(s). The use of topics makes it efficient to find logs by matching topics.

\begin{figure}[b]
\vspace{-0.15cm}
\setlength{\belowcaptionskip}{-0.3cm}
\setlength{\abovecaptionskip}{-0.01cm}
\centerline{\includegraphics[width=1\linewidth]{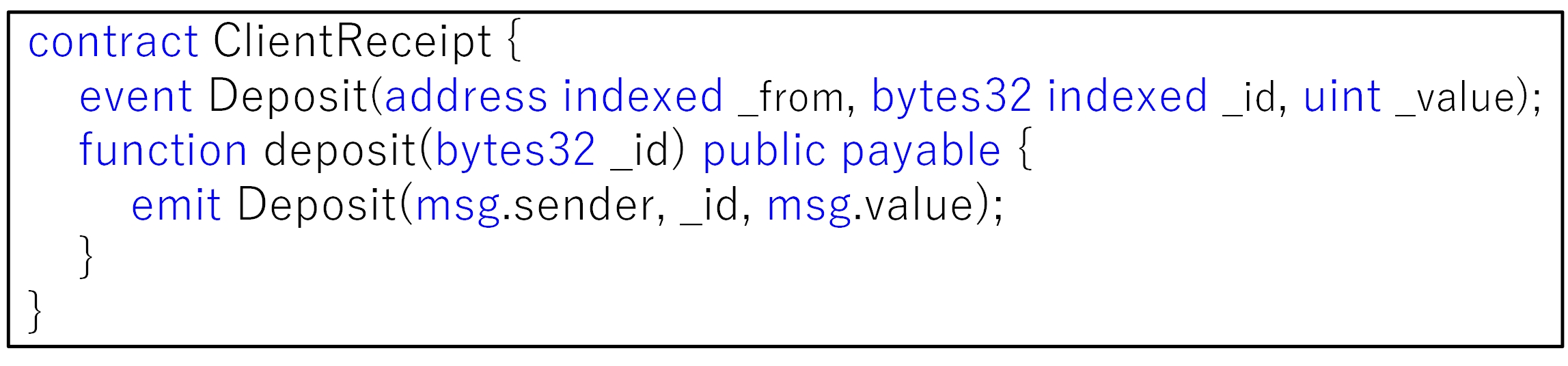}}
\caption{Example of Event Definition and Use.}
\label{example}
\end{figure}

To declare a Solidity event, we begin with the keyword ``event'', then give the identifier, and lastly give a list of typed parameters. In particular, we can add the attribute \texttt{indexed} to certain parameters, which will make them become the topic parts of the log records produced by EVM logging primitives instead of the data parts. In other words, parameters with keyword \texttt{indexed} are searchable. Fig.~\ref{example} gives an example of event definition. For the defined event \texttt{Deposit}, parameters \texttt{\_from} and \texttt{\_id} are searchable. 

To use the defined event, we begin with the ``emit'' keyword, then we give  the name of the event, and finally we provide the arguments (if any) in parentheses. Fig.~\ref{example} also gives an example of the use of the event \texttt{Deposit}. After the event has emitted, {the passed arguments are stored alongside the blockchain to allow retrieving them. More specifically, they are stored in the log entries of the transaction receipt which is produced when a transaction completes.
These logs are connected with the contract address and will be incorporated into the blockchain and stay there provided that a block is accessible.} 
Note that an event generated is not accessible from within contracts, not even the one which has created and emitted them.

Our goal in this paper is to reveal issues with using the event that is defined in a suitable manner, we thus concentrate on studying event use for a certain defined event and hereafter event logging (code) refers to event use (code), and we use them interchangeably. 

\subsection{Solidity Event Applications}
While conventional logging is mainly used for failure diagnosis \cite{SherLog,LogEhance,yuan2012conservative}, the application scenario of Solidity event logging varies. Event logging in principle can have various application scenarios, and the reported typical scenarios include the following \cite{antonopoulos2018mastering}. 

\begin{itemize}[leftmargin=*]
\item \textbf{Execution monitoring to update client front end.} 
{Once events are emitted, they can be listened for by subscribing to catch these events using ethers.js---a complete and compact library for Ethereum. As a result, DApps or other light clients can monitor particular events and act accordingly. In particular, the client front end can thus be updated automatically. 
}
\item \textbf{Alternative cheaper storage venue.} Storing data in permanent Storage area is extremely expensive in Ethereum, dominating the gas cost of a typical transaction. Storing the data in log record instead is much cheaper.
Thus, event logging can serve as an alternative cheaper storage venue. 

\item \textbf{Debugging during development.} { 
Events can be used to record specific activities or states that may occur during the execution of a smart contract, and thus provide an invaluable source of information into the operation of a smart contract. By regularly tracking and auditing the log entries (generated after events have emitted) of the transaction receipt, developers can check whether the smart contract is running as expected and quickly identify any potential bugs or security issues that may arise. 
}
\end{itemize}

\section{Methodology}

\subsection{Subjects and Data Used in Our study}
We use popular Solidity projects hosted on GitHub as the study subjects. { Borges and Valente \cite{BORGES2018112} found through a survey with 400 Stack Overflow users that the stargazers counts are viewed by practitioners as the most useful measure of popularity on GitHub. We thus focus on Solidity projects hosted on GitHub with relatively larger stargazers counts. Smaller stargazers counts represent lower popularity and the corresponding projects are more likely written by novice developers, and studying these projects will less likely reveal current event logging practices by Solidity developers. } As Solidity is a relatively new programming language, we find that the stargazers counts for Solidity projects are small compared to that for other older languages such as Java and C++. Finally, we retrieve all non-fork, non-private 2,915 Solidity repositories whose stargazers counts are larger than 5 (at the time of our data collection date--September 1, 2022) as the study subjects. 
The threshold of 5-star is chosen as follows. {We initially crawled 10,000 Solidity repositories hosted on GitHub and randomly sampled 200 repositories to study their features by checking the accompanied \texttt{project descriptions} and \texttt{Readme} files. The first author conducts the check and reports 18 sample projects (\emph{e.g.}, teach how to write your first solidity program) among the 200 repositories, and the second author confirms that the reported 18 repositories truly are sample projects. The stargazers counts for the 18 projects are all less than or equal to 5. We thus conjecture that for those projects whose stargazers counts are less than or equal to 5, quite a few are sample projects and 
there exists serious homogeneity among them. } 

\begin{figure}[t]
\vspace{-0.15cm}
\setlength{\belowcaptionskip}{-0.3cm}
\setlength{\abovecaptionskip}{-0.01cm}
\centerline{\includegraphics[width=0.9\linewidth]{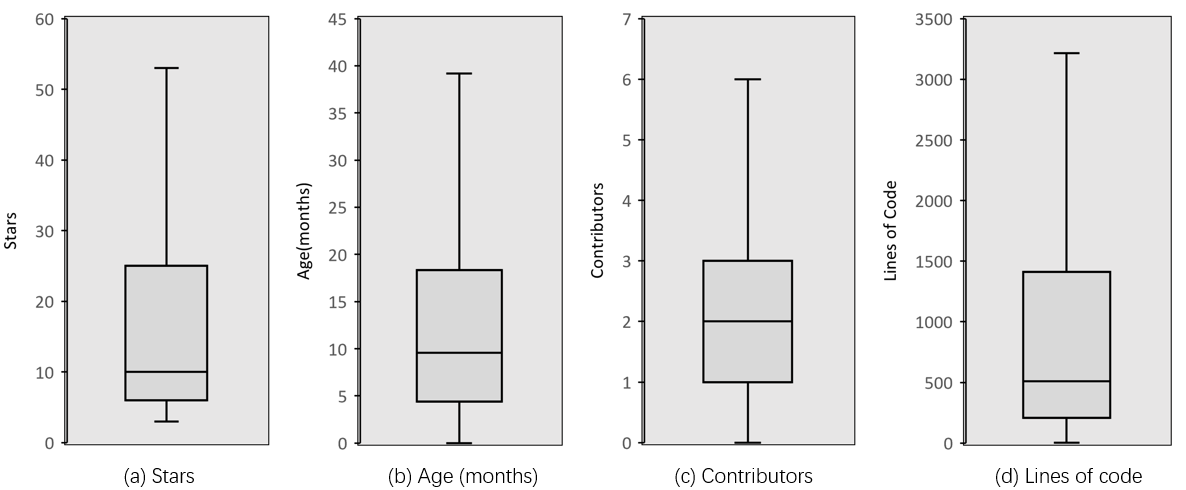}}
    \setlength{\belowcaptionskip}{-0.01cm}
\caption{Boxplots for Some Metrics about the Dataset.}
\label{fig:stastics}
\vspace{-2.15mm}
\end{figure}

{\autoref{fig:stastics} shows boxplots with the distribution of stargazers counts, the age (in number of months), number of contributors, and number of lines of code (LOC, exclude the comments and empty lines) for the 2,915 projects in the dataset (outliers are omitted). The average value is respectively 42.1, 14.4, 3.8, and 1460.1 for stargazers counts, age, number of contributors, and number of lines of code. While the code line numbers of the studied projects are typically smaller than that of popular projects written in other languages, Solidity projects of the studied size are complex and might take much time and effort to audit \cite{grayfuzzing}.
} 

To study the churn rate and evolution of event logging code, we need to explore the evolution history. {As the ages of the considered projects are relatively young (see Figure2b), we account for the full evolution history for each project. In particular, we use the REST API to get all non-merged commits for each project, and further the author name, commit log, and involved files for each commit. }

\subsection{Study Methodology}
We investigate numerous aspects of the practices of event logging. To investigate the density of event logging code, we count the LOC of the event logging code as well as the LOC of the whole program. {To achieve this purpose, we develop a small utility which only counts the actual lines of source code and excludes the comments and empty lines. To recognize the event logging code, we make use of the regular expression ``\textbackslash s+emit\textbackslash s+(\textbackslash w+)\textbackslash s*\textbackslash (([\textbackslash s\textbackslash S]*?)\textbackslash )\textbackslash s*''.}

To study the goodness of event logging practice, we first contrast the churn rate of event logging code with that of the entire code. Code churn is a metric that indicates how often a given piece of code gets edited. If a given piece of code receives changes too often, that's usually a bad sign. The code churn rate is calculated as \textit{Churned LOC}/\textit{Total LOC} \cite{icse05codechurn}, where \textit{Churned LOC} consists of added, modified, or deleted code lines. Likewise, we measure churn rate for event logging code as \textit{Churned Event Use}/\textit{Total Event Use}, where \textit{Churned Event Use} consists of event use lines added, modified, or deleted. {Our developed utility parses each revision in the considered commit history to establish the churned code lines and event use lines, and then the churn rate is calculated accordingly.} The calculated churn rates for all revisions of a certain project are finally averaged to get the final churn rate for this project.

We further establish the percentage of independent event logging code modifications. 
Within the studied commit history, there are in total 43, 550 event logging code modifications. Within these modifications, there exist two kinds: some are solely consistency updates along with other non-event logging code modifications (\emph{e.g.}, bug fix) within the same revision, and others are modifications that modify the event logging code as after-thoughts. 
{The same modification categories have been employed to investigate conventional logging code changes \cite{chen2017characterizing} and annotation changes \cite{yuannotation}. To distinguish these two kinds, one conservative policy is to examine whether the revisions only include changes solely to event use code but not to other non-event use code.} Using this policy, we can get 1,132 event use code modifications. These event use code modifications are absolutely independent event use code modifications, and account for 2.59\% of all event use code modifications. 

However, the policy above is too conservative as developers are inclined to batch multiple code changes into one revision. During our analysis of the event logging code modification, we made one observation: if the variables involved in event logging code modification are not affected by other code changes within the same revision, then this event logging code modification is likely to be independent of other code changes. For instance, given the event logging code modification $\texttt{emit Send(msg.sender, id1)} \rightarrow \texttt{emit Send(msg.sender, id2)}$ where \texttt{Send} is the name of an event and \texttt{id1} and \texttt{id2} are identifiers of two variables, the changed variable is \texttt{id1} and this modification is likely to be an independent modification if \texttt{id1} is not affected by other code changes within the same revision. This makes sense because if the values of the involved variables have not changed, then the reason for event use code change is very likely that we have not appropriately used the event logging feature at first. To determine whether a variable is affected, we conservatively check whether it is redefined or passed as an argument of a function call within the revision. 

On top of this observation, we can get 5,033 independent event logging code modifications, which account for 11.56\% of all {event logging code modifications}. As the result based on this observation can possibly be inaccurate, we further randomly sample 500 modifications and manually verify our analysis results on them {to see the accuracy of our analysis}. This manual verification step is done by two distinct authors, and we deem a checked modification as true independent event logging code modification only when both of the authors have agreed with it. The manual verification suggests that the accuracy of our analysis is 92\%. The Cohen's kappa coefficient is 0.85 for this manual verification, indicating a good agreement rate. There are majorly two sources for the false positive: (1) The event logging code modification is on the string text description, and the text description is related with changes of other non-logging code; (2) The event logging code is dependent on a condition, and the modifications on the condition result in the event logging code modification.

To understand the underlying reasons for independent event logging code modifications, we randomly sample 419 modifications from the 1,132 absolute independent event logging code modifications, which achieves 99\% confidence level and ±5\% confidence interval, and investigate their details. For these sampled modifications, we examine developers’ commit messages about the modifications, relevant source code, code comment, and together with the specific event logging code modifications to understand the modifications. To minimize our own subjective judgment, each modification is checked independently by two authors of the paper. If they cannot clearly understand the reason or have disagreements with the reason for some modifications, we always conservatively classify them into the “unknown” category when our results are presented. The Cohen's kappa coefficient is 0.75 for this manual check, suggesting the agreement rate is good. 

{The above two manual studies in particular consist of three phases. In phase one, the first two authors of the paper independently derive an initial list of the categories (whether or not a modification is independent one for the first study and the reason behind the independent modification for the second study). In phase two, the two authors compare the derived categories and discuss the disagreements, and make refinements to their established categories according to the discussion results. In phase three, the Cohen's kappa coefficient is calculated based on the refined categories. Similar process has been employed by other works \cite{testannotation,overflowcomment}. 

\section{RQ1: Pervasiveness of Event Logging} 

To begin with, we want to see whether event logging feature is indeed widely used by Solidity developers in practice. It is observed that there are 95,803 event uses in total for the latest versions of the studied 2,915 projects in this paper. \autoref{fig:UseByProject} displays the density plot of the event use numbers for the 2,915 projects. We can see from the figure that a majority of projects make use of a moderate number of event logging features, but a couple of projects make use of an exceptionally large number of event logging features. In particular, the average number of event logging uses per project is 32.8. On the whole, it can be concluded that the use of event logging feature is pervasive in Solidity programming by developers.

\begin{figure}
\vspace{-0.15cm}
\centering
\begin{subfigure}{.24\textwidth}
  \centering
  \includegraphics[width=1.0\linewidth]{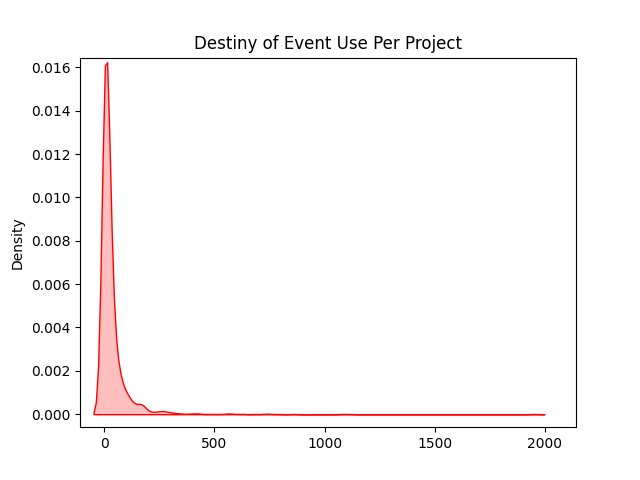}
  \setlength{\belowcaptionskip}{-0.1cm}
\setlength{\abovecaptionskip}{-0.4cm}
  \caption{Event Use by Project}
  \label{fig:UseByProject}
\end{subfigure}%
\begin{subfigure}{.24\textwidth}
  \centering
  \includegraphics[width=1.0\linewidth]{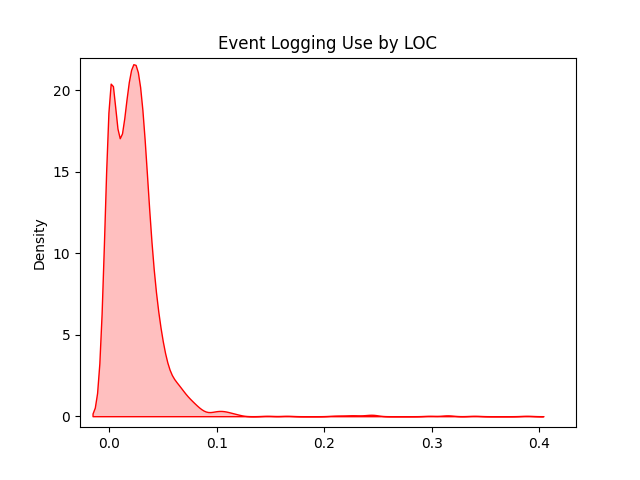}
  \setlength{\belowcaptionskip}{-0.1cm}
\setlength{\abovecaptionskip}{-0.4cm}
  \caption{Event Use by LOC}
  \label{fig:UseByFile}
\end{subfigure}
\setlength{\abovecaptionskip}{-0.01cm}
    \setlength{\belowcaptionskip}{-0.01cm}
\caption{Density of Event Use by Project and LOC.}
\label{fig:UseProjectFile}
\vspace{-2.15mm}
\end{figure}

A potential concern that will arise from \autoref{fig:UseByProject} is whether developers will overuse event logging feature in practice.
{By overuse, we here hypothesize that event logging feature is used frequently with the value of event use per LOC larger than 0.1 and is used in a way that consumes unnecessary more gas.}
To check this, for each studied project, we additionally calculate the value of event use per LOC. The density plot of the calculated value across all the considered 2,915 projects is shown in \autoref{fig:UseByFile}. We can see that the value is small for majority of projects, yet can be greater than 0.1 or even close to 0.4 for certain projects. The minimum, 1st quartile, median, 3rd quartile, and maximum value is 0.0, 0.006, 0.020, 0.031, and 0.388 respectively, and the average value is 0.022. In particular, among the 2,915 projects, the value is larger than 0.1 for 33 projects. If the value is greater than 0.1, it suggests that there is an event use for every less than 10 lines of code. From these data, we can see that while majority of the projects have a reasonable value of event use per LOC, a few projects have an exceptionally large value of event use per LOC and thus potentially suffer from the problem of ``overuse of event logging feature''. 

We further manually check the 33 projects for which the value of event use per LOC is larger than 0.1. We find that a major reason for the high value is that developers use several separate events to record transaction-related information,
while the general case is that developers use a single event (or at least fewer events) to record the information. As an illustration, it can be that the specific event \texttt{event0 (x,y,z)} to record transaction-related information is replaced by three separate events: \texttt{event1(x)}, \texttt{event1(y)}, and \texttt{event1(z)}. The recorded information are essentially the same for these two event use cases, but using more events will cost more gas. Thus, ``overuse of event logging'' raises the potential question of whether the corresponding event definitions are appropriate and (if not) whether refactorings are needed.

\begin{center}
\fcolorbox{black}{gray!25}{\parbox{0.97\linewidth}{
\noindent\textbf{Finding 1}: The average value of event use per project and per LOC is 32.8 and 0.022 respectively, yet the maximal values can be exceptionally large for both of the two metrics. For projects with exceptionally high value of event use per LOC, one major reason is that developers use several separate events instead of a single event to record transaction-related information.

\noindent\textbf{Implication 1}: Solidity event logging feature is used pervasively in practice, but can be overused occasionally. IDE can warn developers if event logging feature is overused, and the numerical data summarized here and detailed online can be referred to set the thresholds. {In addition, to save gas when the event is used, it is preferable that developers define event in a way that contains more parameters. These parameters correspond to different aspects of a certain subject to be logged (\emph{e.g.}, the different addresses involved in a transaction).}}}
\end{center}

\section{RQ2: Goodness of Event Use Practice}

We first want to check whether event use code is actively maintained. Among all the commits in the studied commit history, we find that 11.83\% of them involve modifications to event use code. This suggests that despite the relatively small density of event use code compared to the entire code, event use code takes an exceptionally important part of code evolution. We also compare the code churn rate of the event use code with that of the entire code (as detailed in Section 3.2), and \autoref{fig:ownershipuserate} shows the result. The average value of code churn rate for entire code and event use code is 0.0030 and 0.00282 respectively. Thus, there is no significant difference between these two code churn rates. Overall, we see that event use code has been actively and continuously modified, suggesting that Solidity developers are actively maintaining event use code for software functionality like they maintain other code. 

\begin{center}
\fcolorbox{black}{gray!25}{\parbox{0.97\linewidth}{
\noindent\textbf{Finding 2}: Event use code is modified in a significant number (11.83\%) of all the committed revisions, and its average churn rate is nearly the same as that of the entire code.

\noindent\textbf{Implication 2}: Event use code is being actively and continuously maintained by developers, and it takes an exceptionally important part of code evolution despite its relatively small density.}}
\end{center}

As detailed in Section 3.2, some event use code modifications are dependent ones and others are independent ones. {For all the event use code modifications, we establish a percentage 2.59\% of absolute independent event use code modifications using the conservative policy. Moreover, we identify a percentage 11.56\% of independent event use code modifications using an automatic analysis procedure on top of one observation. Given the accuracy of the automatic analysis is 92\%, it can be deemed that the automatic analysis identifies a percentage 10.64\% (11.56\% $\times$ 92\% = 10.64\%) of independent event use code modifications. Considering that other independent event use code modifications can be neglected by our automatic analysis, it can be deemed that an absolute lower bound 2.59\% and a relaxed lower bound 10.64\% of independent event use code modifications have been established.} Overall, there exists a non-neglectable percentage (with lower bound being 10.64\%) of independent event use code modifications. For these modifications, event use code is not written right by developers at their first attempt and developers later detect the problem and modify the event logging code accordingly. The large percentage of this kind of independent event use code modifications suggests that the current developer practice of using Solidity event feature is not good enough, and some developers are inclined to use event logging feature in an arbitrary and subjective way. After detecting the event use problems, Solidity developers take time and efforts to address them as after-thoughts. {By ``good enough'', we here mean that if developers use event logging feature in a principled and correct way, nearly all of the changes to the event logging code ideally should be consistent updates. In other words, besides consistent updates, no (or few) modifications are needed on the event use code. Thus, use of Solidity event logging feature ``good enough'' will make the percentage of independent event logging modifications very low. In particular, we deem the use of Solidity event logging feature as ``good enough'' if the percentage of independent event logging modifications is below 5.0\%.}

\begin{figure}[t]
\vspace{-0.15cm}
\setlength{\belowcaptionskip}{-0.3cm}
\setlength{\abovecaptionskip}{-0.01cm}
\centerline{\includegraphics[width=0.6\linewidth]{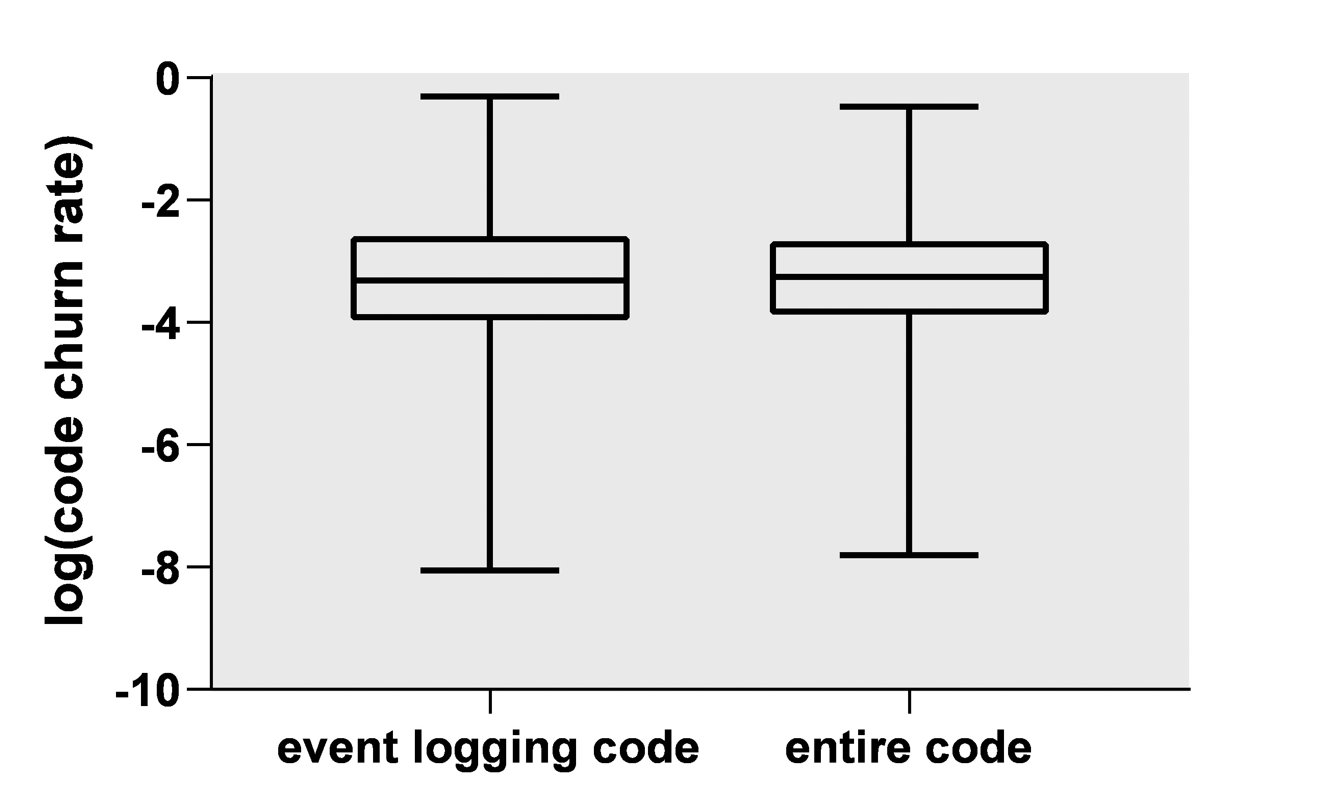}}
    \setlength{\belowcaptionskip}{-0.01cm}
\caption{Code Churn Rate for Event Use and Entire Code.}
\label{fig:ownershipuserate}
\end{figure}

\begin{center}
\fcolorbox{black}{gray!25}{\parbox{0.97\linewidth}{
\noindent\textbf{Finding 3}: There exists a significant percentage (with lower bound being 10.64\%) of event use code modifications that are independent of other non-event logging code changes.

\noindent\textbf{Implication 3}: The current practice of using event logging feature is not good enough, {introducing issues to the quality of event logging that developers will address as after-thoughts}. Tools that support the systematic testing of event logging behaviours would be helpful. { Besides, the Solidity event logging feature should be more extensively discussed
in books and tutorials, so that developers can make better-informed implementation decisions.
}
}}
\end{center}

\section{RQ3: Reasons for Event Logging Code Modifications}
In this section, we give the results of our manual analysis of the randomly sampled 419 independent event use code modifications.

Overall, the 419 modifications can be divided into 5 categories as shown in \autoref{tab1}. The particular meanings of the 5 categories are as follows: the category \emph{Parameter Change} means that the parameters of event use code have changed but the name of the used event does not change, the category \emph{Addition} refers to adding an event use, the category \emph{Deletion} refers to the deletion of an event use, the category \emph{Move} means that the location of the event use code has been moved but its parameters and name remain the same, and finally the category \emph{Replacement} means that the name of the used event is modified without modifying the parameters. For three modifications, they change both the name and parameters of the used event, so we do not classify them into any category. 

We can see from the table that the instances for categories \emph{Parameter Change}, \emph{Addition}, and \emph{Deletion} are most dominating, accounting for 92.36\% of all sampled modifications. Having tools to help check for problematic event parameters ({i.e., event parameters that should be substituted by other parameters for better logging quality}), missing use of events, and unnecessary use of events can thus help developers better use events the first time they write code, reducing subsequent event use changes. {In particular, the high percentage of \emph{Parameter Change} and \emph{Addition} instances is in line with existing studies on conventional logging, and the explored automatic logging support for ``what to log'' and ``where to log'' is also invaluable for Solidity event logging \cite{learntolog, LogEhance}. } By contrast, the instances for categories
\emph{Move} and \emph{Replacement} are relatively limited. 

\begin{center}
\fcolorbox{black}{gray!25}{\parbox{0.97\linewidth}{
\noindent\textbf{Finding 4}: Within independent event use code modifications, developers spend significant efforts in changing parameters, adding a new event use, and deleting an existing event use (accounting for 92.36\% of all modifications). They seldom move an event use or replace an event use with another one.

\noindent\textbf{Implication 4}: Tool builders should focus on developing tools for helping developers to detect and fix issues related with problematic event parameters, missing event use, and unnecessary event use. {The explored automatic logging support about ``what to log'' and ``where to log'' for conventional logging is also invaluable for Solidity event logging, and a viable solution is using the ``learn to log'' principle on top of high quality training data. 
}
}}
\end{center}

\begin{table}[t]
      \setlength{\abovecaptionskip}{0.1cm}
    \setlength{\belowcaptionskip}{-0.1cm}
      \small
\caption{Category of Independent Event Use Code Changes}
\begin{center}
\begin{tabular}{|c|c|c|}
\hline
\textbf{Change type} & \textbf{\#Instance} & \textbf{\#Percentage}\\
\hline
Parameter Change & 168 & 40.09\%\\
Addition & 145 & 34.61\%\\
Deletion & 74 & 17.66\%\\
Move & 24 & 5.73\%\\
Replacement & 5 & 1.19\%\\
\hline
\end{tabular}
\label{tab1}
\end{center}
\end{table}

We next give a detailed analysis of the reasons for the modifications in the 5 categories. 

\subsection{Parameter Change}
Category \emph{Parameter Change} is most common according to our data (with a percentage of 40.09\%), and its specific reasons are also the most diverse.

\subsubsection{Unintended Wrong Usage}
41.67\% (70 instances) of modifications in this category arise because unintended wrong value related with a certain non-address variable has been recorded by event use. The wrong value emerges due to typical programming errors related with variables. Among the 70 instances, there are 25, 12, 9, and 3 instances that fall into change kind \emph{replace a variable with another variable that is in scope}, \emph{arithmetic error change}, \emph{replace a wrapped method call (for a variable) with another method call that is similar in identifier name}, and \emph{wrap/unwrap a variable with a method call} respectively. As the search space for these change kinds is relatively small, there exists the prospect of well-designed tools to automatically conduct these changes. Given much progress has been made in recent years towards automatically detecting \cite{multiple-fault,guifault,yufse, flsurvey, yuguifl} and fixing \cite{6035728, monperrus2018automatic, yuEmSE, yutse, yuseip} relatively simple bugs, this prospect is huge. An example is shown in Fig.~\ref{Parameter}(a) where the parameter \texttt{referrer} is wrapped with method call \texttt{plasmaOf} in the new version. According to the definition of \texttt{plasmaOf}, depending on a condition, the return value of it may or may not be the argument passed in (the \texttt{referrer} in this case). According to the change, what needs to be logged is the return value of the function (for an argument) but not the argument itself, so the direct use of argument \texttt{referrer} is unintended and wrong. 

\subsubsection{Mixed Use of Addresses}
20.24\% (34 instances) of modifications in this category are related with mixed use of addresses. {It is a common practice to use event to record transaction-related information in Solidity programming, and the information in general will include the addresses involved in the transaction (i.e., the addresses of the recipient and sender).} However, when using event to record information, addresses are often mixed, and the most common one is mixing the sender’s address with that of the caller. For example, given a scenario where contract \texttt{A} calls contract \texttt{B} to transfer \texttt{X} tokens in account \texttt{C} to account \texttt{D}, then account \texttt{C} is the actual token sender and account \texttt{D} is the actual token recipient. In event logging, we then should record the addresses of account \texttt{C} and account {D}. However, a mistake that many developers are prone to making is to record \texttt{msg.sender} as the actual token sender, that is, record \texttt{A} as the actual token sender. In other words, the address for caller of the contract is mixed with that for the actual token transferer. Fig.~\ref{Parameter}(b) gives a real example where \texttt{src} is the actual token transferer. However, developer mistakenly records the contract caller \texttt{msg.sender} as the token transferer.

\subsubsection{Improving String Description}
9.52\% (16 instances) of modifications in this category are improving string description, making the content in the string easier to understand or closer to what the developer wants to express. Besides logging value, event logging feature can also be used to log one or several string descriptions for explaining what value has been logged or the purpose of the logged value. However, developers may not pay enough attention to the string description at the beginning, and they gradually improve it. An example is shown in Fig.~\ref{Parameter}(c) where the initial string description ``\texttt{\_mint b}'' is much less informative than the new string description ``\texttt{initOptinoToken}''. 

\subsubsection{Gas Saving}
6.55\% (11 instances) of modifications in this category are related with gas saving, which replaces Storage type variable with Memory type variable to save gas. The reason is after compiling the event use to EVM bytecode, an extra \texttt{Sload} EVM operation would be needed to access the variable if it is Storage type, and the \texttt{Sload} operation costs 800 gas since it deals with data in Storage area. Thus, when we use event to store the value of a certain variable in transaction log, local Memory type variable would be preferable to Storage type variable if they hold the same value. Fig.~\ref{Parameter}(d) gives an example, the original variable \texttt{delay} is a global variable and the new variable \texttt{delay\_} is a function parameter, which will respectively be Storage type and Memory type in Solidity. Since their values are the same for event \texttt{DelaySet}, using Memory type variable \texttt{delay\_} can save gas. {Given the utmost relevance of reducing the gas costs of smart contracts in the blockchain ecosystem \cite{optimizationtosem}, it would be extremely worthwhile to investigate the gas impact of this kind of variable replacement on the whole Ethereum net in the future. }

\vspace{1.0mm}
For the remaining 37 (22.02\%) instances in this category, we are not clear or have not agreed with the underlying reasons. 

\begin{center}
\fcolorbox{black}{gray!25}{\parbox{0.97\linewidth}{
\noindent\textbf{Finding 5}: For changes that change parameters of event use, four key reasons are: unintended wrong variable-related value has been recorded (41.67\%), mixing addresses of transaction sender, transaction recipient, and contract caller (20.24\%), improving string description (9.52\%), and gas saving (6.55\%). For wrong variable-related value, the wrong value emerges due to typical programming errors related with variables, such as arithmetic error and forgetting to wrap the variable with a method call. For gas saving, developers replace Storage type variable with Memory type variable that has the same value.

\noindent\textbf{Implication 5}: Testing event logging behaviours is important and testing efforts should pay special attention to typical programming errors related with variables and the addresses involved in event use. With regard to wrong variable-related value, there exists the prospect of well-designed tools to automatically conduct changes to correct the value as the change space is relatively small. With regard to gas saving, tools can automatically help developers conduct the change from Storage type variable to Memory type Variable and compilers can do this gas optimization during the optimization process. {Besides, developers should take the string description in event use seriously, making it as informative as possible.
}}}
\end{center}

\begin{figure}[b]
\centerline{\includegraphics[width=0.90\linewidth]{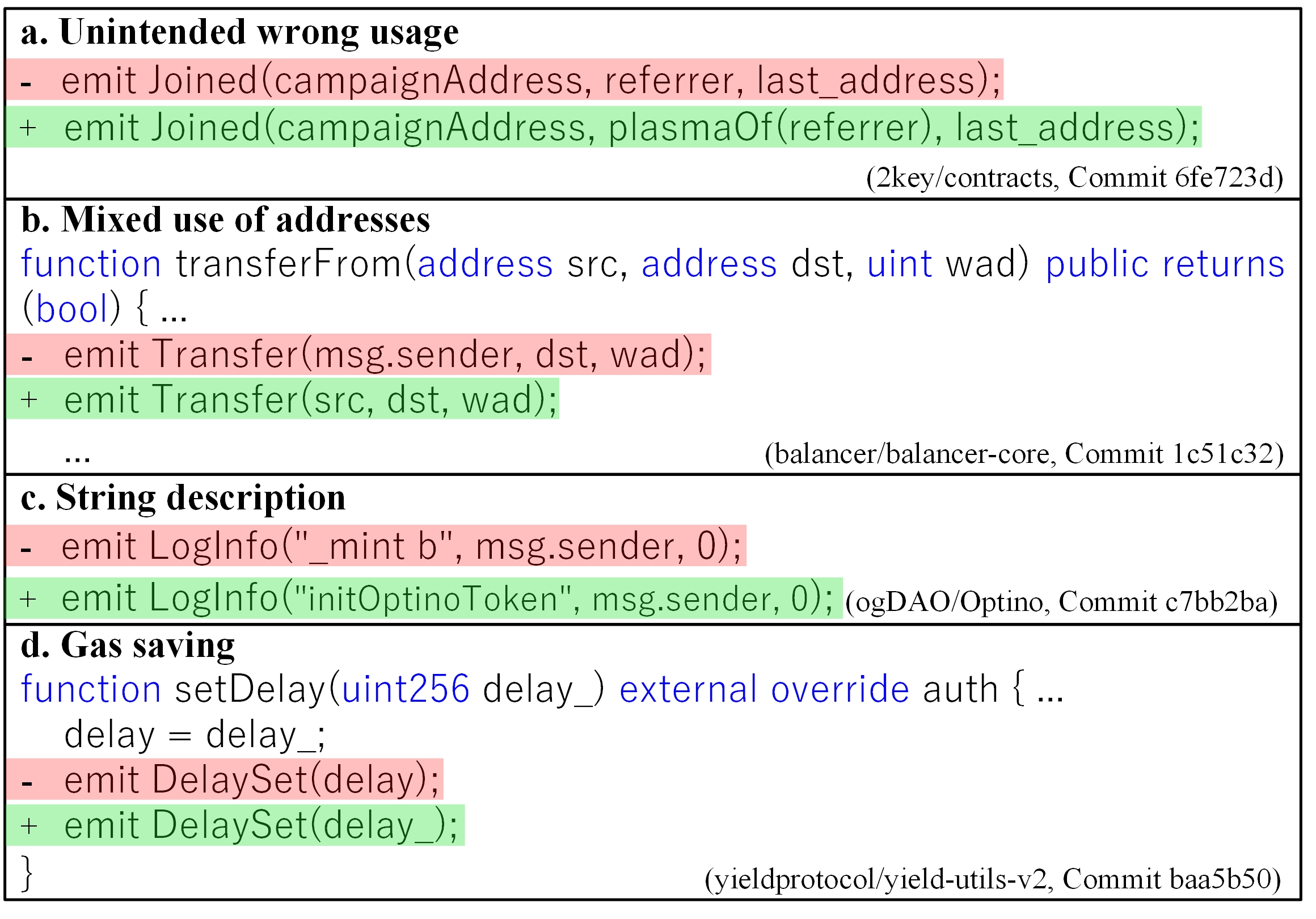}}   \setlength{\abovecaptionskip}{-0.01cm}
    \setlength{\belowcaptionskip}{-0.01cm}
\caption{Example of Parameter Change.}
\label{Parameter}
\end{figure}

\subsection{Addition}
Category \emph{Addition} accounts for a significant percentage (34.61\%) of the 419 modifications, and the underlying reasons are as follows.

\begin{figure}[htbp]
\centerline{\includegraphics[width=0.90\linewidth]{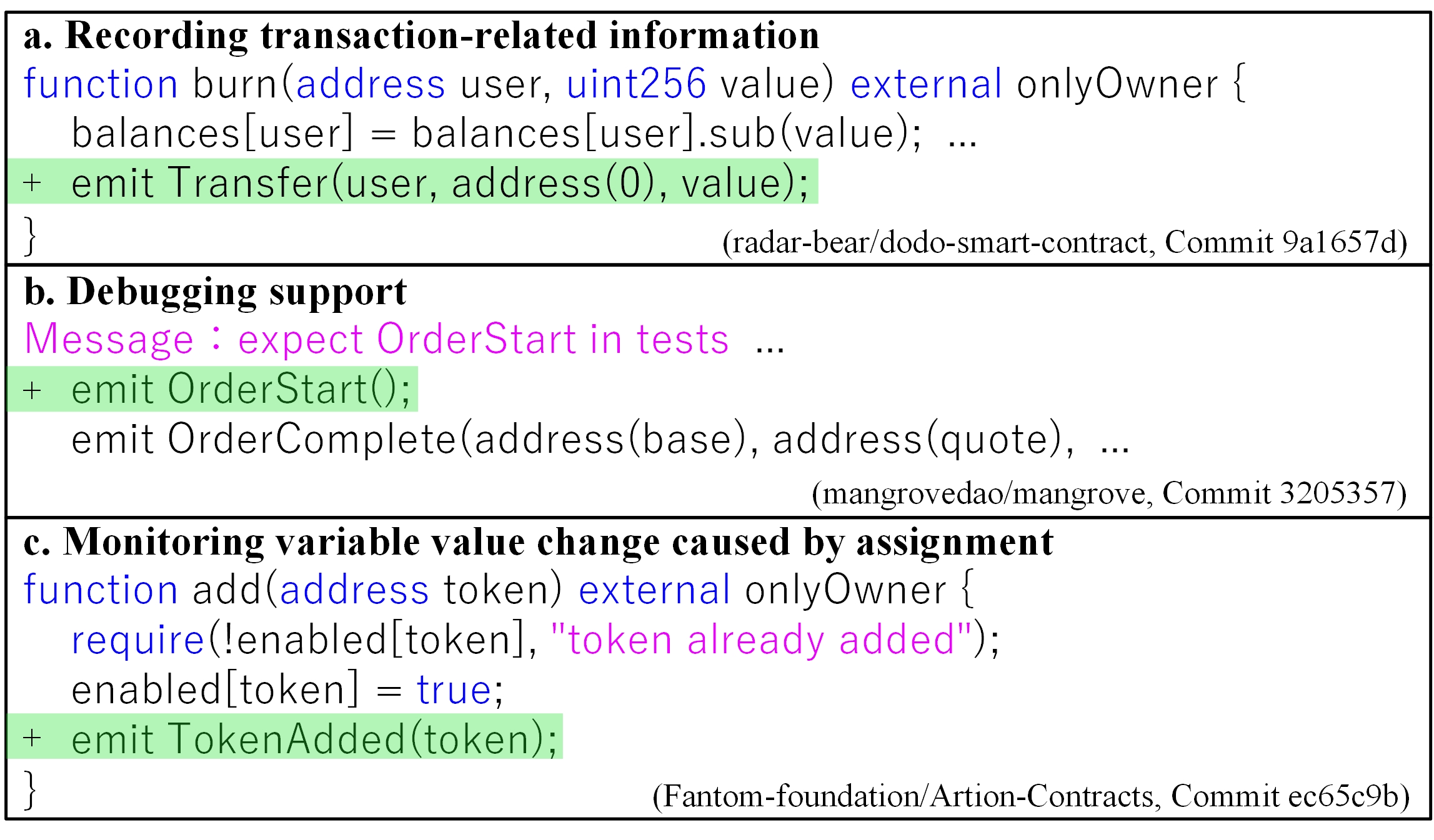}}
\setlength{\abovecaptionskip}{-0.01cm}
    \setlength{\belowcaptionskip}{-0.01cm}
\caption{Example of Addition Change.}
\label{Addition}
\end{figure}

\subsubsection{{Recording Transaction-related Information}}
35.86\% (52 instances) of modifications in this category are post-transaction additions, which add an event use after the asset transfer code to record transaction-related information. The information typically include the addresses involved in the transaction, the amount, the balance, etc. If event logging is employed to record transaction-related information, the Ethereum platform will notify the client of the occurrence of the transaction. As a result, there is no need to continuously check the Ethereum ledger to determine whether the transaction exists. 
An example is shown in Fig.~\ref{Addition}(a). The newly added \texttt{Transfer} event records the addresses \texttt{user} and \texttt{address(0)} of the two parties involved in the transaction, as well as the number of transferred tokens \texttt{value}.

\subsubsection{Debugging Support} 
18.62\% (27 instances) of modifications in this category are related with facilitating debugging. Solidity does not provide developers with language-level debugging facilities typically found in other languages (\emph{e.g.}, \texttt{print}). The difficulty of debugging is thus increased, and developers can use event logging to facilitate debugging. There are various ways in which developers can use event logging to help with debugging, including logging the value of an intermediate variable, logging string information related to debugging, using empty events without parameters (note in this case, transaction logs will still have information related with event such as its identifier), etc. An example is shown in Fig.~\ref{Addition}(b) where \texttt{OrderStart} is an empty event, and the (trimmed) commit message confirms that this modification is for debugging.

\subsubsection{{Monitoring Variable Value Change Caused by Assignment.}}
17.93\% (26 instances) of modifications in this category are post-assignment additions, where the assignment includes not only the direct variable-to-variable assignment, but also the assignment of results of typical operations (such as function call and math operation, but excluding transaction operation) to variables or elements of more complex data structure such as array. Event logging in this case can typically be used to monitor the value change for variables of interest, updating front end of light clents and DApp services accordingly. An example is shown in Fig.~\ref{Addition}(c). After using an assignment ``\texttt{enabled[token]=true}'' to change the value of \texttt{enabled[token]}, the event use ``\texttt{TokenAdded(token)}'' is employed to indicate that now the account with address \texttt{token} is allowed to participate in something such as a transaction. 

\vspace{1.0mm}
The use of event logging after transaction and assignment operations is of great significance, in particular for DApps. The high percentage of post-transaction and post-assignment additions suggests that developers frequently forget to use event logging after transaction and assignment operations, and the commit messages for most of these modifications clearly explain that the developers forget to use events at the beginning. Thus, dedicated tools could offer support for recommendations of event uses for these operations. 
For the remaining 40 (27.59\%) instances in this category, we are not clear or have not agreed with the underlying reasons. 

\begin{center}
\fcolorbox{black}{gray!25}{\parbox{0.97\linewidth}{
\noindent\textbf{Finding 6}: For changes that add an event use, three major reasons are: recording transaction-related information (by post-transaction additions, 35.86\%), debugging support (18.62\%), and monitoring variable value change caused by assignment (by post-assignment additions, 17.93\%).

\noindent\textbf{Implication 6}: {Research on providing automatic ``where to log'' support could focus on transaction and assignment operations, and more systematic work is needed for exploring in detail the specific types of transaction and assignment operations for which event logging is highly desired. Besides, it is worthwhile to further explore the attributes of code snippets for which event logging is added to support debugging. }
 }}
\end{center}

\subsection{Deletion}
Category \emph{Deletion} accounts for a large percentage (17.66\%) of the 419 modifications, and the underlying reasons are as follows.

\begin{figure}[b]
\centerline{\includegraphics[width=0.90\linewidth]{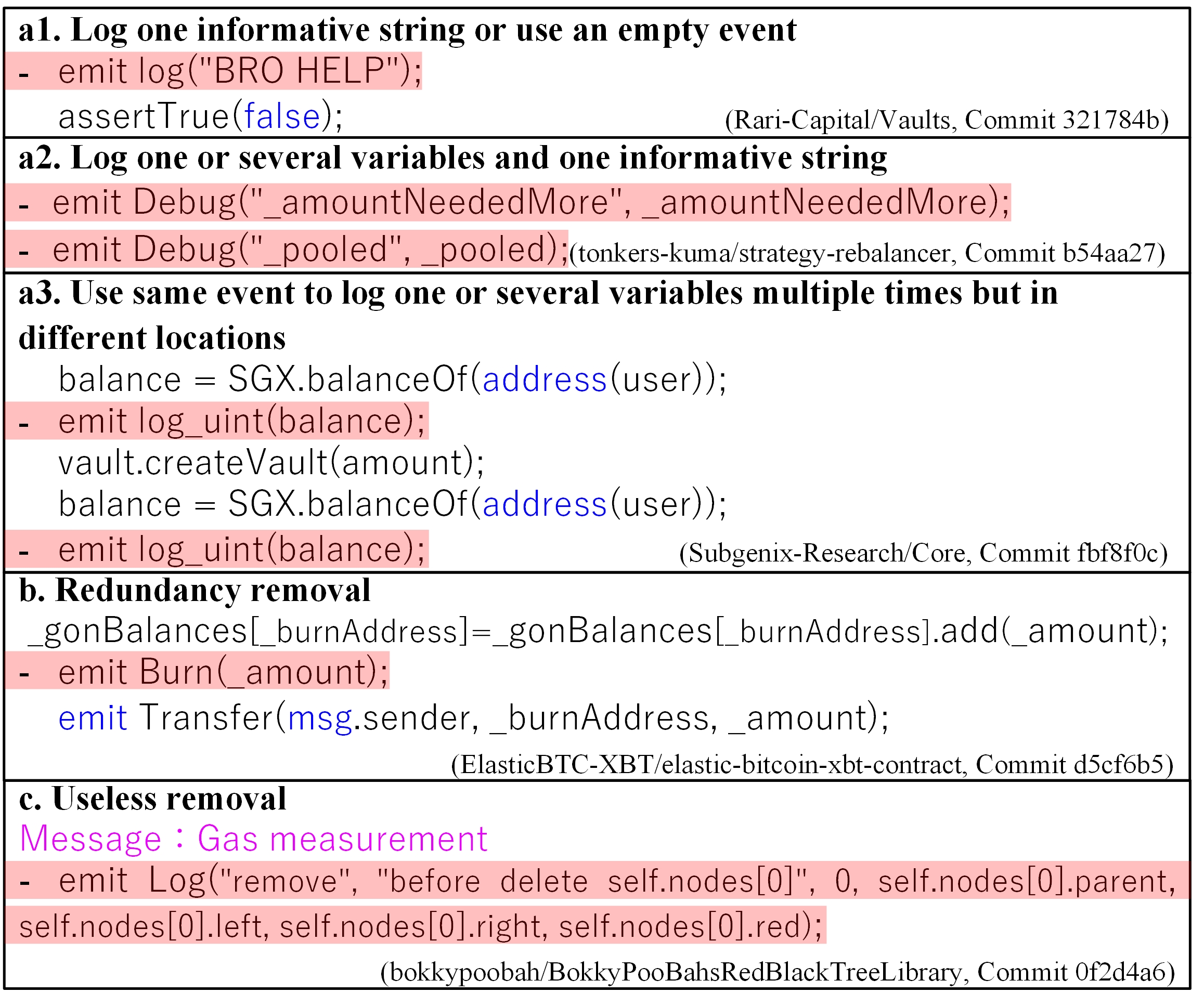}}
\setlength{\abovecaptionskip}{-0.01cm}
    \setlength{\belowcaptionskip}{-0.01cm}
\caption{Example of Deletion Change.}
\label{remove}
\end{figure}

\subsubsection{Debugging Removal}
58.11\% (43 instances) of modifications in this category are related with removing event logging previously used for debugging. As these event debugging removals are independent of other non-logging code changes, so these event debugging removals in principle can actually also be done in last commit or earlier commits (more specifically at the code version for which the use of a certain event for code debugging has finished). 
Thus, the high percentage of modifications in this category suggests that developers often do not delete events that are no longer needed in a timely manner after using event feature for debugging, negatively impacting code readability and increasing gas cost. However, it is typically impossible to analyze whether the developer has finished debugging from the code only. {The first author checks the 43 event debugging removals, and finds that there are three common forms of event logging debugging. The second author confirms the result, and the three forms are as follows: }

\begin{itemize}[leftmargin=*]
\item \textbf{Log one informative string or use an empty event.} In this case, developer uses event logging to log a string description that may have special meaning for himself or use an empty event as exemplified in category \emph{Addition}. Fig.~\ref{remove}(a1) gives an example of this form. 

\item \textbf{Log one or several variables and one informative string.} In this case, developer uses event logging to log the value of one or several variables, and one informative string that is typically related with the variable name(s). The purpose is to judge whether the variable value is consistent with the expected value. Fig.~\ref{remove}(a2) gives an example of this form. 

\item \textbf{Use same event to log one or several variables multiple times but in different locations.} In this case, developer uses same event multiple times to see whether the value change of a certain variable is expected, facilitating debugging. Fig.~\ref{remove}(a3) gives an example of this form. 

\end{itemize}

Given these common forms of event logging debugging, tools can check them and inform developers to delete them in case debugging has already finished. 

\subsubsection{Redundancy Removal}
12.16\% (9 instances) of modifications in this category are removing redundant event logging code. That is, if the recorded content by a certain event use is the same as or already included in the recorded contents by some other event uses, then the event use can be deleted to save gas.  Fig.~\ref{remove}(b) gives an example of this change type. Since the value of the parameter \texttt{\_amount} is logged by both events \texttt{Burn} and \texttt{Transfer}, one of them is deleted.

\subsubsection{Useless Removal}
There are two additional instances (2.70\%) in this category that arise for removing useless event logging code. If event logging records something that is unnecessary or not used at all, then it should be deleted since EVM logging primitives cost gas. In general, it is hard to determine whether the developer has used useless event logging code from the source code only. We need to also consider the commit message if available. Fig.~\ref{remove}(c) gives an example where the commit message clearly says the purpose of the deletion.

\vspace{1.0mm}
{Note that the underlying reasons for the above 54 change instances can be viewed more generally as gas saving in a way}. For the remaining 20 (27.03\%) instances in this category, we are not clear or have not agreed with the underlying reasons. {In particular, we are not aware of any deletion instances of ``dead'' event logging code in our sample. It can be that such change instances will nearly always be consistency updates, and further work needs to be conducted to investigate this problem. }

\begin{center}
\fcolorbox{black}{gray!25}{\parbox{0.97\linewidth}{
\noindent\textbf{Finding 7}: For changes that delete an event use, two major reasons are: remove no-longer needed event logging previously used for debugging (58.11\%) and remove redundant event logging (12.16\%). The event logging used for debugging typically has three forms: log one informative string or use an empty event, log one or several variables and one informative
string, and use same event to log variable(s) multiple
times in different locations.

\noindent\textbf{Implication 7}: Developers often do not delete event uses that are no longer needed in a timely manner. Based on the common forms of event logging debugging, dedicated tools can check them and remind developers to delete them if debugging has finished. Also, {by conducting advanced analysis (\emph{e.g.}, alias analysis),} dedicated tools can check whether there exist redundant event uses that record the same content and suggest the redundancy removal.}}
\end{center}

\subsection{Move}
Category \emph{move} accounts for a relatively small percentage (5.73\%) of the 419 modifications. The reasons for changes in this category cannot be easily determined. We list the types of move below and try to give reasons for some of them. 

\begin{figure}[htbp]
\centerline{\includegraphics[width=0.90\linewidth]{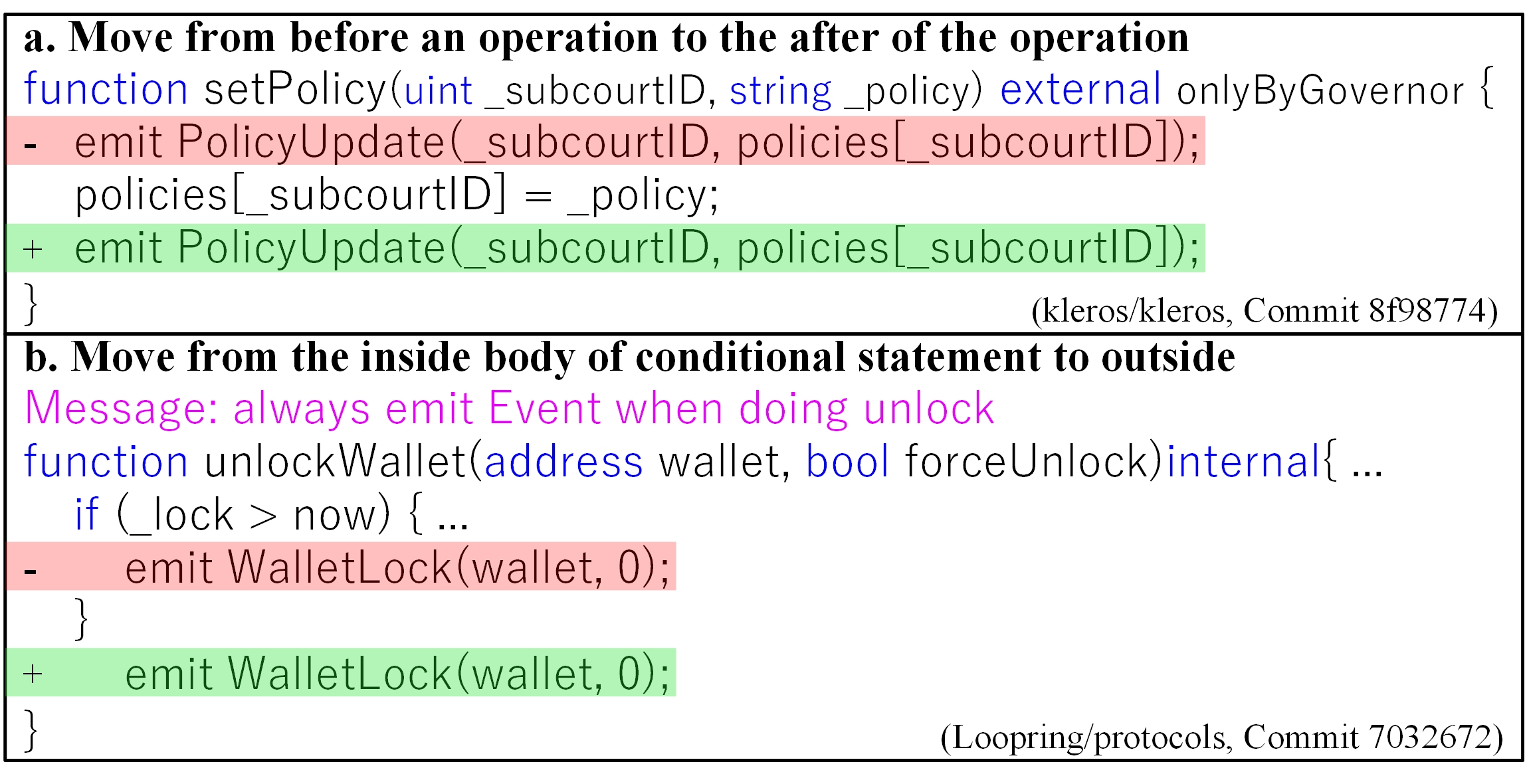}}
\setlength{\abovecaptionskip}{-0.01cm}
    \setlength{\belowcaptionskip}{-0.01cm}
\caption{Example of Move Change.}
\label{move}
\end{figure}

58.33\% (14 instances) of modifications in this category are moving the event use code from before an operation (such as transaction, assignment, authorization, etc.) to the after of the operation. 
An example is shown in Fig.~\ref{move}(a).
This change type is in line with common sense, that is, ``\emph{finish things first and then record the completed things}''. It is also compatible with the significant number of post-transaction and post-assignment additions of event use in Category \emph{Addition}. The high percentage of this type of changes suggests that it is desirable for IDE to warn developers if they put event logging code before an operation. 

29.17\% (7 instances) of modifications in this category are moving the event use code from the inside body of conditional statement or loop statement to outside, or vice versa. 
An example is shown in Fig.~\ref{move}(b). We can see from the commit message that developer has changed the conditions for the event logging to trigger. For the other 6 instances of this type, the commit messages do not have a clear explanation for the change, but we surmise that the reason for this type of change is either the condition for triggering event logging has changed or the original triggering condition is wrong. 

There are 2 change instances in this category which instead move the event logging code from after an operation to the before of the operation. This change type is anti-intuition and relatively rare, and there are usually special reasons behind it. For one change instance (
commit 33de75b), the developer's commit message is ``\emph{This triggered a warning when static analysis tools were used, because emitting an event is considered a state change.}", so the change aims to avoid  triggering warnings from static analysis tools. For the other change instance (
commit 3cc7b64), the developer's commit message is ``\emph{Moved the CreatePool event before the AddLiquidity event for cleaner subgraph code}", so the purpose of the change it to clean code. Due to the small number of occurrences, there is little commonality in the reasons.

\begin{center}
\fcolorbox{black}{gray!25}{\parbox{0.97\linewidth}{
\noindent\textbf{Finding 8}: For changes that move an event use, developers frequently move an event use from before an operation (such as transaction, assignment, authorization, etc.) to the after of the operation (58.33\%). In addition, they also often  move the event use from the inside body of conditional statement or loop statement to outside, or vice versa (29.17\%).

\noindent\textbf{Implication 8}: {Developers are conservative in moving event use code}. Given event use in general is placed after an operation, dedicated tools can check whether event use is placed before an operation and warn developers about this {abnormal} behaviour.}}
\end{center}

\subsection{Replacement}
Category \emph{Replacement} accounts for a very small percentage (1.19\%) of the 419 modifications (note for changes in this category, even though the recorded value has not changed, but the names of the two events are different and this will be reflected in transaction logs), and it arises as the wrong event is used at the beginning or there is an event that is more suitable for recording the needed information. With regard to the reason, it is speculated that developer defines too many events or the locations of the event definitions are scattered, so they forget which events have been defined. It is in general impossible to determine whether a certain event use should be replaced by another event use, but we can justify some of these modifications by analyzing whether the name of the function where the event use is located has a corresponding relationship with the name of the event.
An example is shown in Fig.~\ref{Replacement}. In this example, the developer uses the \texttt{EscrowWithdrawn} event in the \texttt{refund} function, which is later changed to the \texttt{EscrowRefunded} event. It can be seen that the name of the event used at the beginning does not match with the name of the function where it is located. 

\begin{figure}[htbp]
\centerline{\includegraphics[width=0.90\linewidth]{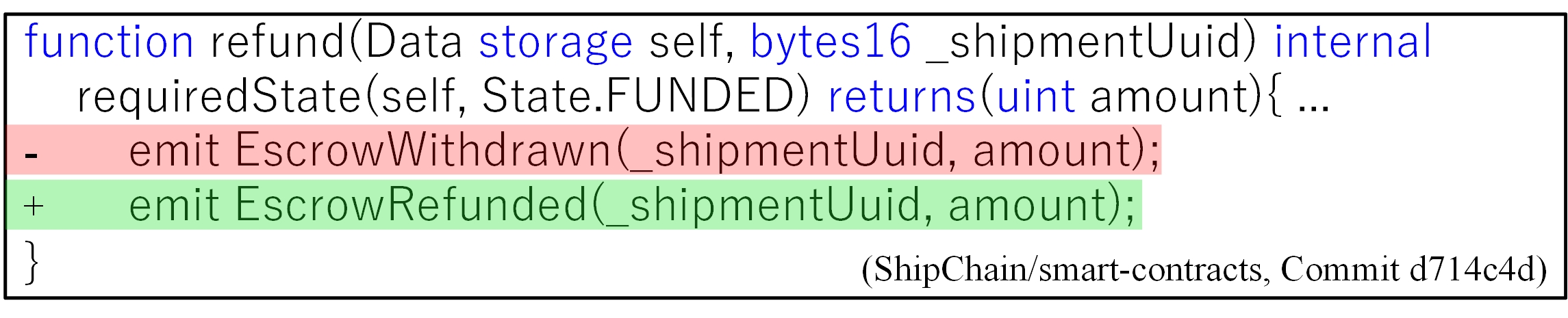}}
\setlength{\abovecaptionskip}{-0.01cm}
    \setlength{\belowcaptionskip}{-0.01cm}
\caption{Example of Replacement Change.}
\label{Replacement}
\end{figure}

\vspace{-0.7cm}
\section{Problematic Event Logging Parameter Checker for Gas Saving}
To demonstrate the possibility of automatic event logging support on top of our findings, {we design a simple parameter-level checker to detect a particular type of problematic event logging code which consumes extra gas than needed}. The checker is motivated by the non-neglectable number of parameter changes that replace Storage type variable with Memory type variable (Finding 5), {and contributes to saving gas.}

Our observation is that when we use event logging to store the value of a certain variable in transaction log, Memory type variable would be preferable to Storage type variable if they hold the same value. Note that both the Storage type variable and Memory type variable are already declared elsewhere in the code, we are not changing their definitions and we just reference them as event parameters. Since these two variables share the same value, the stored value in the transaction log will be the same. Since we are not changing the definitions and value of Storage type variables, the persistence of Storage type variable on the blockchain will not be impacted. 
Thus, changing the parameter of an event use from a Storage type variable to another Memory type variable (when these two variables share the same value) will not impact code semantics. 

{The checker is implemented in approximately 800 lines of Python excluding dependencies, comments, and empty lines, and is built upon 
\texttt{python-solidity-parser} \cite{github} which can partially compile Solidity code snippet. In particular, the checker relies on the parsed AST to work.} It
takes the path of the contract or the project directory as an input, and then analyzes whether each contract in the directory has gas optimization opportunities and outputs the following information if so: (1) The contract line number for which the gas can be optimized; (2) The specific event use in the line to optimize; (3) The specific Storage type variable of the identified event use;
(4) The specific Memory type variable to replace the identified Storage type variable. In particular, the checker contains two major phases. First, it identifies all event use code in the contract whose parameters involve a certain Storage type variable. Then, it checks whether there exists an in-scope Memory type variable whose value is the same as the Storage type variable. 

{To investigate the effectiveness of the tool on the most popular projects, we applied the checker to the top 200 (in terms of stargazers counts) GitHub Solidity projects}. The checker detected that 35 projects suffered from this issue for their latest versions of code. In total, there are 207 problematic event logging uses. {For each of the 35 projects, we opened an issue which includes the following information: (1) A short message explaining the purpose of our study; (2) The explanation of our finding about gas saving opportunities involving Memory and Storage type variables; (3) The exact code line(s) and variable(s) for which gas can be optimized based on our finding (with the code snippet embed); (4) A question about confirmation of the detected problem.} The owners of 9 projects have already confirmed the detected problems (confirming 103 problematic event logging uses in total) and most of them have merged the changes into the code base. {We have received very positive feedback from Solidity developers. One developer said ``This is interesting, thanks a lot for taking the time to identify this optimization''. } This result suggests that on top of our findings, even a simple, proof-of-concept checker can effectively contribute to improving the quality of Solidity event logging instructions. In particular, the result confirms that to obtain systematic and automatic assistance for better event logging, the first essential step is to understand the current manual efforts involved with event logging. 

\section{Threats to Validity}

\noindent
\textbf{External Validity.}
First, our study uses Solidity projects hosted on GitHub and these projects may not be sufficiently diverse or representative. {Thus, whether the results can be generalized to other Solidity projects (\emph{e.g.}, hosted on BitBucket) is a potential issue.} To mitigate this threat, {our study accounts for a considerable number of popular Solidity projects on GitHub and the existing study \cite{grayfuzzing} suggests that popular Solidity projects on GitHub will cover diverse application domains (like auctions and tokens)}. 
{Second, we sample certain independent event logging code modifications to understand the underlying reason. To reduce the sampling bias, we have ensured that the results fall under the confidence level of 95\% with a confidence interval of ±5\%. Third, while we use stargazers counts to select popular projects and previous works have suggested the existence of many popular, deployed projects on GitHub \cite{grayfuzzing, durieux2020empirical}, it is possible that sample, not-deployed projects sneak into our dataset. Finally, the proof-of-concept checker is evaluated only on the top 200 popular GitHub projects and more evaluations need to be conducted to see its effectiveness in a more general sense.}

\noindent
\textbf{Internal Validity.} {Our study methodology in Section 3.2 involves two manual processes and bias can possibly be introduced. To mitigate this threat, each manual analysis will involve two distinct authors of the paper. They independently conduct the analysis and the analysis results will be compared.} In particular, for the manual examination of the characteristics of the 419 sampled independent event logging code modifications, we examine developers’ commit messages, source code, together with the event logging code modifications to reason about the modifications.  

\noindent
\textbf{Construct Validity.} {For all of our developed programs, in particular the utility to calculate LOC and the proof-of-concept checker, we have performed thorough testing to ensure their correctness. }

\section{RELATED WORK}
\noindent
\textbf{Empirical Study on Smart Contract.}
The transaction-reverting statement can effectively protect smart contracts against abnormal or malicious attacks. To explore how developers can enhance transaction-reverting statements, Liu et al. \cite{liu2021characterizing} analyze the security implications of transaction-reverting statements.  
Mariano et al.  \cite{mariano2020demystifying} conduct an empirical study of loops in Ethereum smart contracts and they cluster smart contract loops according to their semantic features. 
Durieux et al. \cite{durieux2020empirical} conduct an empirical study to analyze and compare several state-of-the-art smart contract analysis tools.
{To our knowledge, our work is the first to explore Solidity event logging feature and gives a systematic study about density and in particular evolution of Solidity event logging code.}

\noindent
\textbf{Smart Contract Analysis and Verification.}
As vulnerabilities in smart contract can potentially be exploited, recent years have witnessed a surge of proposed approaches to ensure that smart contracts are free of vulnerabilities. Securify \cite{10.1145/3243734.3243780} and Ethainter \cite{10.1145/3385412.3385990} leverage the rewriting system Datalog to detect vulnerabilities through pattern matching. Luu et al. \cite{application3} present an approach which uses symbolic execution to find potential security bugs in smart contracts. Rodler et al. \cite{rodler2021evmpatch} design a framework for on-the-fly automatic patching of faulty smart contracts.
Ren et al. \cite{ren2021making} design a security-enhanced code suggestion module on top of a bidirectional LSTM network. 
{Compared to these approaches, our work instead aims to uncover the practical issues faced by Solidity developers (when using the event feature) and inspire the development of analysis and verification methods to alleviate the issues.}

\noindent
\textbf{Research on Logging.}
In general, logging messages can improve program comprehension and reduce maintenance costs. Yuan et al. \cite{logging,SherLog} conduct a series of studies about logging in system software, covering characteristics study, tool design for log enhancement, etc. 
Chen and Jiang \cite{chen2020studying} conduct the first large-scale empirical study of the use of Java log tools in the wild. 
Kim et al. \cite{kim2020automatic} propose a log analysis system on top of log history. The system uses statistical text mining techniques to calculate significance and noise scores for each log line, and then highlights abnormal log lines based on the calculated scores. Le et al. \cite{le2021log} propose a log-based anomaly detection method that does not demand parse of log messages. 
{Solidity event logging features a few important distinctions compared to conventional logging (as given in Section 1), and thus it is important to study how Solidity developers use event logging in practice. We in this paper fill the research gap by conducting a large-scale empirical study about Solidity event logging practices in the wild.
} 

\section{CONCLUSION}
We provide the first quantitative characteristic study of the current Solidity event logging practices using 2,915 Solidity projects hosted on GitHub. The study methodically investigates the pervasiveness of event logging, the goodness of event logging practice, and in particular the reasons for event logging code evolution, and delivers 8 original and important findings. We additionally give the implications of our findings, and these implications are beneficial for researchers, language designers, tool builders, and developers in order to improve all facets of Solidity event logging. Based on one of our findings, we develop a proof-of-concept checker and effectively detect problematic event logging code that consumes extra gas in 35 popular GitHub projects and 9 project owners have confirmed the detected issues. {For future work, we would like to develop tools that conduct systematic testing of event logging behaviours based on our findings. We are also passionate about building an advanced tool on top of our findings which can effectively support automatic code transforms related with Solidity event logging, including bug fixing, refactoring, and optimization.}

\section{Data Availability}
Our replication package (including code, dataset, etc.) is available at \url{https://github.com/zhongxingyu/Solidity-Event-Study}.

\section*{Acknowledgments}
\noindent
We are grateful to the anonymous reviewers for their insightful comments. This work was partially supported by National Natural Science Foundation of China (Grant No. 62102233), Shandong Province Overseas Outstanding Youth Fund (Grant No. 2022HWYQ-043), and Qilu Young Scholar Program of Shandong University. 

\bibliographystyle{ACM-Reference-Format}
\balance
\bibliography{sample-base}

\end{document}